\input harvmac
\newcount\figno
\figno=0
\def\fig#1#2#3{
\par\begingroup\parindent=0pt\leftskip=1cm\rightskip=1cm\parindent=0pt
\baselineskip=11pt
\global\advance\figno by 1
\midinsert
\epsfxsize=#3
\centerline{\epsfbox{#2}}
\vskip 12pt
{\bf Fig. \the\figno:} #1\par
\end insert\endgroup\par
}
\def\figlabel#1{\xdef#1{\the\figno}}
\def\encadremath#1{\vbox{\hrule\hbox{\vrule\kern8pt\vbox{\kern8pt
\hbox{$\displaystyle #1$}\kern8pt}
\kern8pt\vrule}\hrule}}

\overfullrule=0pt

%
\def\tilde{\widetilde}
\def\bar{\overline}
\def\Z{{\bf Z}}
\def\T{{\bf T}}
\def\S{{\bf S}}
\def\R{{\bf R}}

\font\zfont = cmss10 

\def\bigone{\hbox{1\kern -.23em {\rm l}}}
\def\ZZ{\hbox{\zfont Z\kern-.4emZ}}

\Title{hep-th/9602070, IASSNS-HEP-96-08}
{\vbox{\centerline{STRONG COUPLING EXPANSION}
\bigskip\centerline{ OF CALABI-YAU COMPACTIFICATION  
}}}
\smallskip
\centerline{Edward Witten\foot{Research supported in part
by NSF  Grant PHY92-45317.}}
\smallskip
\centerline{\it School of Natural Sciences, Institute for Advanced Study}
\centerline{\it Olden Lane, Princeton, NJ 08540, USA}\bigskip

\medskip

\noindent

In a certain strong coupling limit, compactification of the 
$E_8\times E_8$ heterotic string on a Calabi-Yau manifold $X$
can be described by  an eleven-dimensional theory compactified on
$X\times \S^1/\Z_2$.  In this limit, the usual relations among low 
energy gauge couplings hold, but the usual (problematic) prediction 
for Newton's constant does not.
In this paper, the equations for unbroken supersymmetry are expanded
to the first non-trivial order, near this limit, verifying the consistency
of the description and showing how, in some cases, if one tries
to make Newton's constant too small, strong coupling develops in one
of the two $E_8$'s.  The lower bound on Newton's constant 
(beyond which strong coupling develops) is estimated
and is relatively close to the actual value.

\Date{February, 1996}
\newsec{Introduction}

This paper will consist mainly of a technical calculation
in eleven dimensions, which could be of conceivable relevance to
grand unified model building.  To explain this, I will begin with an
extended introduction focussing on the
 model building, postponing most of the eleven-dimensional
details to section two.

The models of particle physics that are most straightforwardly
derived from the $E_8\times E_8$ heterotic string -- with an assumption
of small string coupling constant at all scales and
low energy  $N=1$ supersymmetry in four dimensions -- are in many
ways strikingly attractive.  
For example, one naturally gets
the right gauge groups and fermion quantum 
numbers and recovers the usual, successful super-GUT prediction
for low energy gauge couplings, even though
 other aspects 
of the GUT package are substantially modified
(for instance, the usual and not obviously desireable
GUT relations among Yukawa couplings do
not arise; there are generally \ref\wen{X.-G. Wen and E. Witten,
``Electric And Magnetic Charges In Superstring Models,''
Nucl. Phys. {\bf B261} (1985) 651.}
unconfined particles with fractional electric charge with GUT masses
but not fitting in GUT multiplets).

\nref\ginsparg{P. Ginsparg, 
"Gauge and Gravitational Couplings in Four-Dimensional 
  String Theories", Phys. Lett. B197 (1987) 139.}
\nref\kap{
V.S. Kaplunovsky, "One Loop Threshold Effects In String Unification",
  Nucl. Phys. B307 (1988) 145,  erratum  Nucl. Phys. B382 (1992) 436.}
\nref\nan{ I. Antoniadis, J. Ellis, R. Lacaze, and D.V. Nanopoulos, 
       "String Threshold Corrections and Flipped SU(5)", 
        Phys. Lett. B268 (1991) 188; S. Kalara, J. L. Lopez, and
D. V. Nanopoulos, ``Threshold Corrections And Modular Invariance In
Free Fermionic Strings,'' Phys. Lett. {\bf B269} (1991) 84.} 
\nref\mayr{P. Mayr, H.P. Nilles, and S. Stieberger, "String Unification and 
        Threshold Corrections", Phys. Lett. B317 (1993) 53.}
A major difficulty of the framework (as exhibited in early
computations of the unification scale \refs{\ginsparg - \mayr}) 
has been the following.
In contrast to usual GUT models, which do not incorporate gravity
and so make no prediction for Newton's constant,  these perturbative
string models
do make a definite prediction for the gravitational coupling strength, a 
prediction that comes out too large.  If expressed as a prediction
for Newton's constant, the error is  typically a factor of 400; if expressed
(as it naturally arises) as a prediction for the logarithm of the Planck
mass, the error is about 6\% (but given the accuracy with which the
low energy gauge couplings are now known, the discrepancy
is six or seven standard deviations).
Various proposals have been put forward
for dealing with this problem in the context of
perturbative string theory, but none of them is obviously compelling.
For discussions of some of the possible  scenarios, which
include large threshold corrections, extra matter fields,
higher or exotic Kac-Moody
levels, GUT models more strictly embedded in strings, and anisotropic
Calabi-Yau manifolds,  see
\nref\lew{ D.C. Lewellen, "Embedding Higher-Level Kac-Moody Algebras
       in Heterotic String Models", Nucl. Phys. B337 (1990) 61.}
\nref\afont{    A. Font, L.E. Ibanez, and F. Quevedo, "Higher-Level Kac-Moody
       String Models and Their Phenomenological Implications",
       Nucl. Phys. B345 (1990) 389.}
\nref\ibanez{ G. Aldazabal, A. Font, L.E. Ibanez, A.M. Uranga,
      "String GUT's", Nucl. Phys. B452 (1995) 3,
``Building GUTS From Strings,'' hep-th/9508033.} 
\nref\chaud{S. Chaudhuri, S.-W. Chung, G. Hockney, and J. Lykken,
 ``String Consistency For Unified Model Building'';
S. Chaudhuri, G. Hockney, and J. Lykken, ``Three Generations In The
Fermionic Construction.''}
\nref\finnel{D. Finnell, 
``Grand Unification with Three Generations in Free Fermionic String Models,''
hep-th/9508073.}
\nref\ib{ L.E. Ibanez, "Gauge Coupling Unification:  Strings
      vs. SUSY GUT's", Phys. Lett. B318 (1993) 73.}
\nref\march{K. R. Dienes, A. E. Faraggi, and J. March-Russell,
``String Unification, Higher-Level Gauge Symmetries, 
and Exotic Hypercharge Normalizations,'' hep-th/9510223.} 
\nref\dix{ L.J. Dixon, V.S. Kaplunovsky, and J. Louis, "Moduli Dependence of 
       String Loop Corrections to Gauge Coupling Constants", 
       Nucl. Phys. B355 (1991) 649.}   
\nref\udienes{K. R. Dienes
and A. E. Faraggi, ``Making Ends Meet: String Unification And Low-Energy
Data,'' Phys. Rev. Lett. {\bf 75} (1995) 2646, ``Gauge Coupling
Unification In Realistic Free-Fermionic String Models,''
Nucl. Phys. {\bf B457} (1995) 409.}
\nref\lopez{J. L. Lopez, D. V. Nanopoulos, and K. Yuan,
``The Search For  A Realistic
Flipped $SU(5)$ Standard Model''; J. L. Lopez and D. V. Nanopoulos, 
``A New Scenario For
String Unification,'' hep-ph/9511426.}
\nref\dienes{K. R. Dienes, ``String Theory And The Path To Unification:
A Review Of Recent Developments,'' hep-th/9602045.}
\refs{\lew - \dienes}; the last reference contains an extensive review
and references.

As an alternative to approaches that depend on weak coupling,
one might wonder whether the problem has a natural solution in
a region of {\it large} string coupling constant.  Certainly when the
string coupling constant is large, there might be large corrections
to the predicted value of Newton's constant, but at the cost of possibly
ruining the successful weak coupling predictions for gauge couplings.  
General arguments have been given \ref\banks{T. Banks
and M. Dine, ``Coping With Strongly Coupled String Theory,''
Phys. Rev. {\bf D50} (1994) 7454; M. Dine,  ``Coming To Terms With
Strongly Coupled Strings,'' hep-th/9508085.}
for why there might exist a strong coupling region in which the usual
predictions would hold for gauge couplings but not for Newton's constant,
but these arguments require some assumptions about wave function
renormalization.  

With the understanding of the strong coupling behavior of string theory
that has been obtained of late, it seems appropriate to revisit this
question: Is there a regime of strongly coupled string theory in which
the four-dimensional gauge couplings are small and obey the usual GUT
relations, but Newton's constant
is significantly smaller than usual?  The answer, as we will see,
is ``yes''; the desired properties hold in  a
regime in which the ten-dimensional
string coupling constant is large, and the volume of the Calabi-Yau is
also large, in such a proportion that 
the four-dimensional effective gauge couplings are small.

\nref\witten{E. Witten, ``String Theory Dynamics In Various Dimensions,''
Nucl. Phys. {\bf B433} (1995) 85.}
\nref\dab{A. Dabholkar, ``Ten Dimensional Heterotic String As A Soliton,''
Phys. Lett. {\bf B357} (1995) 307.}
\nref\hull{C. M. Hull, ``String-String Duality In Ten Dimensions,''
Phys. Lett. {\bf B357} (1995) 545.}
\nref\polwitten{J. Polchinski and E. Witten, ``Evidence For Heterotic
- Type I String Duality,'' hep-th/9510169, to appear in 
 Nucl. Phys. {\bf B460} (1996) 525.}
To analyze this region, first note that if the Calabi-Yau volume goes
to infinity fast enough compared to the string coupling  constant,
the strong coupling behavior can be deduced from what happens in
ten dimensions for strong coupling.  Here the behavior is completely
different depending on whether one considers the $SO(32)$ or $E_8\times E_8$
heterotic string.  For $SO(32)$ the strong coupling limit is described
by a weakly coupled ten-dimensional Type I superstring theory 
\refs{\witten - \polwitten}, 
while for  $E_8\times E_8$ the strong coupling limit involves
an eleven-dimensional description \ref\horava{ P. Horava and E. Witten,
``Heterotic And Type I String Dynamics From Eleven Dimensions,''
hep-th/9510209,
to appear in Nucl. Phys. {\bf B460} (1996) 506.}.  In one case,
the gauge fields propagate on the boundary of the world-sheet, and in the
other case they propagate on the boundary of space-time.
In either case, as we will see, in the limit of large Calabi-Yau
volume and large string coupling, one can naturally preserve the usual
predictions for gauge coupling constants while making Newton's constant
smaller.
  
Since neither the length scale of string theory $\sqrt{\alpha'}$
nor the volume $V$ of the Calabi-Yau manifold $X$ (measured in the string
metric) nor the expectation value of the dilaton field $\phi$ is directly
known from experiment, one might think that by adjusting 
$\alpha'$, $V$, and $\langle\phi\rangle$
 one can fit to any desired
values of Newton's constant, the GUT scale $M_{GUT}$, 
and the GUT coupling constant $\alpha_{GUT}$,
thus imitating the situation that prevails in conventional GUT theories.
Let us first recall why things do not work out that way for the weakly
coupled heterotic string.  In ten dimensions, the low energy 
effective action looks like\foot{Slightly varying conventions
concerning the weak coupling formulas that appear in the present
paragraph (and will not be used later in the paper) can be found
in the literature.}
\eqn\happy{L_{eff}= -
\int d^{10}x \sqrt g e^{-2\phi}\left({4\over (\alpha')^{4}}R
+ {1\over (\alpha')^{3}}\tr F^2+\dots\right).}
After compactification on a Calabi-Yau manifold $X$ 
of volume $V$ (in the string
metric), one gets a four-dimensional effective action that looks like
\eqn\snappy{L_{eff}=-\int d^4x\sqrt g e^{-2\phi}V
\left({4\over(\alpha')^{4}}R
+{1\over(\alpha')^{3}}\tr F^2+\dots\right).}
The important point is that the same function $Ve^{-2\phi}$ 
 multiplies both $R$ and $\tr F^2$.
Because of $T$-duality one can assume
\eqn\jappy{V\geq (\alpha')^3}
in order of magnitude; for $V\sim (\alpha')^3$ one must use conformal
field theory (and not classical geometry) to compute the effective
$V$ to be used in \happy.  From \snappy, we get at tree level
\eqn\turbo{\eqalign{G_N  &= {e^{2\phi}(\alpha')^4\over 64\pi V}\cr
                   \alpha_{GUT} & ={e^{2\phi}(\alpha')^3\over 16\pi V},
\cr}}
 so that
\eqn\nurbo{G_N= {\alpha_{GUT} \alpha'\over 4}.}
This means that the string scale, controlled by $\alpha'$, is known
in terms of $G_N$ and $\alpha_{GUT}$, and in particular the string
mass scale is not much below the Planck scale.  On the other hand,
for weak coupling the GUT scale cannot be much smaller.
If we suppose that $e^{2\phi}\leq 1$ so that the ten-dimensional string
theory is not strongly coupled, we get 
\eqn\gurbo{
V\leq {  (\alpha')^3 \over \alpha_{GUT}}.}
For a more or less isotropic Calabi-Yau, $V\sim M_{GUT}^{-6}$, and
then the upper bound \gurbo\ on $V$ translates into
\eqn\normo{G_N\geq {\alpha_{GUT}^{4/3}\over M_{GUT}^2},}
which is too large.\foot{Note that the problem might be ameliorated
by considering an anisotropic Calabi-Yau, for instance
one with a scale $\sqrt {\alpha'}$ in $d$ directions and $1/M_{GUT}$ in
$6-d$ directions (with some fairly severe restrictions on $d$ and the
Calabi-Yau manifold $X$ 
to ensure that it is the large dimensions in $X$ that control
the GUT breaking), so that $ V\sim (\alpha ')^{d/2}
/M_{GUT}^{6-d}$.
The amelioration obtained this way, if too small, could possibly
be combined with the strong coupling effect considered below.}

Rather than pursuing one of the possible solutions of this problem
in weak coupling, we will here suppose
that the ten-dimensional string coupling constant is
{\it large}.  There are different strong coupling limits one could 
consider, depending, for instance, on what happens to $V$ while $\phi\to
\infty$.  If $V$ and $\phi$ are both suitably large,
then the theory can be analyzed using a knowledge of the
{\it ten-dimensional} strong coupling behavior.
That is the region we will consider in this paper.

First we consider the  $SO(32)$ heterotic string,
which  involves less novelty because the strong coupling limit
of this theory is simply another string theory -- the Type I superstring
theory in ten dimensions.  We repeat the above discussion, using
the Type I dilaton $\phi_I$, metric $g_I$, and scalar curvature $R_I$.
The analog of \happy\ is 
\eqn\wappy{L_{eff}= -\int d^{10}x \sqrt {g_I} 
\left(e^{-2\phi_I}{4\over (\alpha')^{4}}R_I
+  e^{-\phi_I}{1\over (\alpha')^{3}}\tr F^2+\dots\right).}
The point is that in contrast to the heterotic string, the gravitational
and gauge actions multiply different functions of $\phi_I$, namely
$e^{-2\phi_I}$ and $e^{-\phi_I}$, respectively, since one is
generated by a world-sheet path integral on a sphere and one on a disc.
The analog of \snappy\ is then
\eqn\nappy{L_{eff}=-\int d^4x\sqrt g_I V_I\left({4e^{-2\phi_I}
\over(\alpha')^{4}}R
+{e^{-\phi_I}\over(\alpha')^{3}}\tr F^2+\dots\right)}
($V_I$ is the Calabi-Yau volume measured in the Type I metric),
and the couplings become
\eqn\urbo{\eqalign{G_N  &= {e^{2\phi_I}(\alpha')^4\over 64\pi V_I}\cr
                   \alpha_{GUT} & ={e^{\phi_I}(\alpha')^3\over 16\pi V_I}
                         .\cr}}
Hence
\eqn\kirbo{G_N={e^{\phi_I}\alpha_{GUT}\alpha'\over 4},}
showing that after taking $\alpha_{GUT}$ from experiment and
adjusting $\alpha'$ so that the string scale is comparable to the 
experimentally inferred GUT scale,
one can make $G_N$ as small as one wishes simply by taking $e^{\phi_I}$ to
be small, that is, by taking the Type I superstring to be weakly coupled.
Of course, when the Type I coupling is weak, the usual  tree level relations
among gauge couplings will hold (in a suitable class of string  vacua).

Note that this discussion does not require
a knowledge that the Type I superstring is equivalent to the strong
coupling limit of the $SO(32)$ heterotic string.  We have simply
shown directly that for the weakly coupled Type I superstring the usual
contradiction with measured GUT parameters does not arise (and by the same
token, just as in standard GUTs, 
one gets no prediction for the value of $G_N$ in terms of known
quantities).  This assertion is not essentially new, though it has
not been much used in attempts at string phenomenology.

We will now argue that the $E_8\times E_8$ heterotic string
has an analogous strong coupling behavior: one keeps the standard
GUT relations among the gauge couplings, but loses the prediction
for Newton's constant, which can be considerably smaller than
the weak coupling bound.  There is, however, one important
difference from $SO(32)$. Except under
a certain topological restriction that will be explained, we cannot
make $G_N$ {\it arbitrarily} small in the strongly coupled $E_8\times E_8$
heterotic string (keeping other moduli fixed)
without losing control on the discussion.  There is a lower
bound, whose order of magnitude we will estimate, on how small $G_N$
can be in such a model.

The ten-dimensional $E_8\times E_8$ heterotic string has for its strong
coupling limit $M$-theory on $\R^{10}\times \S^1/\Z_2$ \horava.
The gravitational field propagates in bulk over $\R^{10}\times \S^1/\Z_2$,
while the $E_8\times E_8$ gauge fields propagate only at the $\Z_2$ fixed 
points.
We write $M^{11}$ for $\R^{10}\times \S^1$ and $M^{10}_i$, $i=1,2$ 
for the two components of the fixed point set.
The gauge and gravitational kinetic energies take the form
\eqn\mippy{L=-{1\over 2\kappa^2}\int_{M^{11}}
 d^{11}x \sqrt g R -\sum_i
{1\over 8\pi
(4\pi \kappa^2)^{2/3}}\int_{M^{10}_i}d^{10}x\sqrt g \tr F_i^2.}
A few points about this formula need to be explained.  First,
 $\kappa $ is here
the eleven-dimensional gravitational 
coupling; implicit in \mippy\ is a relation
of the 
ten-dimensional gauge coupling to $\kappa$ that
will be obtained elsewhere \ref\newhorava{P. Horava and E. Witten,
to appear.}. Second,  
as written, the Einstein action in \mippy\ involves an integral
over $M^{11}= \R^{10}\times \S^1$, it being understood that the fields are 
$\Z_2$-invariant; if one wishes the integral to run over $M^{11}/\Z_2$, one
must multiply the gravitational action by a factor of two. Finally, $F_i$, for
$i=1,2$, is the field strength of the $i^{th}$ $E_8$, which propagates
according to \horava\ on the $i^{th}$ component of the fixed point set, 
that is on $M^{10}_i$.

Now compactify to four dimensions 
on a Calabi-Yau manifold $X$ whose volume
(in the eleven-dimensional metric) is $V$.\foot{This is a change
in notation, as earlier $V$ was the Calabi-Yau volume in the
ten-dimensional string metric.  In the remainder of the paper
we use always the metric of the eleven-dimensional theory.}  
Let the $\S^1$ have
radius $\rho$ or circumference $2\pi \rho$.  
Upon reducing \mippy\ to four dimensions, one learns that
 the four-dimensional $G_N$ and $\alpha_{GUT}$ are
\eqn\kippy{\eqalign{G_N & = {\kappa^2\over 16\pi^2 V\rho} \cr
                    \alpha_{GUT} & = {(4\pi \kappa^2)^{2/3}\over 2V}.\cr}}
In so far as experiment shows that $\alpha_{GUT}<<1$, the second formula
shows that the dimensionless number $\kappa^{4/3}/V$ is small.
That is just as well, because the eleven-dimensional description
is only valid under that restriction.  The fact that $\kappa^{4/3}/V$ 
is small also means that the fields propagating on $M^{10}_i$ are weakly
coupled so that, if the vacuum is determined by an $E_8\times E_8$ gauge
bundle of the sort usually used in string theory compactification
(for instance, the standard embedding of the spin connection in the gauge group
or one of the usual generalizations)
then the standard GUT relations among gauge couplings will hold.
After picking $\kappa$ and $V$ to get the experimentally inferred value
of $\alpha_{GUT}$, and also choosing $V$ so that $V^{-1/6}$ agrees with
the experimentally inferred GUT mass scale $M_{GUT}$, 
the first equation in \kippy\
shows that we may also make $G_N$ small by making $\rho$ large.
That is again a favorable result, since for $\rho $ to be large compared
to the eleven-dimensional Planck scale is a necessary condition for validity
of the eleven-dimensional description.

In fact,  the validity of the eleven-dimensional
description certainly requires
 that $\rho$ and $V$ should be large compared to the
eleven-dimensional Planck length and Planck volume.  To the extent that those
are the only restrictions, the above formulas show that while keeping
$\alpha_{GUT}$ and $M_{GUT}$ fixed and making $\rho$ sufficiently big,
we can make $G_N$ as small as we wish.  
That is the actual state of affairs 
under a certain topological restriction, which is roughly that
the vacuum gauge bundle
has the same instanton number in each $E_8$. 
More generally (for instance if one considers
the standard embedding of the spin connection in the gauge group, for 
which the instanton number  vanishes in one of the two $E_8$'s), we will
find that the validity of the eleven-dimensional description  
requires  a further condition which in order of magnitude
is $\rho\leq V^{2/3}/\kappa^{2/3}$.  
(When this does not hold, one is driven to strong coupling in one of
the two $E_8$'s  by a mechanism
analogous to that in \polwitten.)  In order of magnitude,
this translates
into $G_N\geq \kappa^{8/3}/V^{5/3} =\alpha_{GUT}^2V^{1/3}$.  Setting
$V^{1/3}=M_{GUT}^{-2}$, the lower bound on  Newton's constant
becomes in order of magnitude 
\eqn\pillory{G_N\geq {\alpha_{GUT}^2\over M_{GUT}^2}.}
(As in the weak coupling  case discussed in a footnote after equation
\normo, this bound can be made smaller by considering an
anisotropic Calabi-Yau.) 

Comparing this to the bound $G_N\geq \alpha_{GUT}^{4/3}/M_{GUT}^2$ of
weakly coupled string theory, we see that the effect of going to
strong coupling is to make the lower bound on $G_N$ smaller by a factor
of $\alpha_{GUT}^{2/3}$ (or to remove the lower bound entirely if
the instanton numbers are equal in the two $E_8$'s).  
We will evaluate the bound somewhat more precisely in the next section,
and in the approximation considered there, we get the critical
value of Newton's constant (at which strong coupling develops in one of 
the $E_8$'s) to be 
\eqn\ippo{G_N^{crit}={\alpha_{GUT}^2\over 16\pi^2}\left|\int_X
\omega\wedge{\tr F\wedge F-{1\over 2}\tr R\wedge R\over 8\pi^2}\right|,}
with $\omega$ the Kahler form of $X$ and
$F$ the field strength of either of the $E_8$'s.  The
integral on the right hand side of \ippo\ is plausibly
$M_{GUT}^{-2}$ 
times a number of order one. $\alpha_{GUT}$ is this formula is the
coupling in the $E_8$ that is not strongly coupled.
Because of the factor of $16\pi^2$
in the denominator, \ippo\ may be small enough to agree with experiment,
for reasonable values of the integral.  Our calculation in section
two is, however, only carried out in a linearized approximation,
and it is not clear to what extent it might be modified by the
nonlinear terms.  (We will in section three compute the nonlinear terms
for compactifications to six dimensions.)

\bigskip\noindent
{\it Perturbative Expansion }

The rest of this paper will consist 
primarily of a rather detailed calculation
expanding the eleven-dimensional vacuum  to lowest order in $1/V$ and
$1/\rho$.  This mainly involves 
showing how the four-form field strength $G$
of the eleven-dimensional 
theory is turned on without breaking supersymmetry.
A non-zero $G$ field appears in any compactification, as we will see.

By performing this calculation, we will get an interesting test of the 
consistency of the sort of 
vacua considered here and get some intuition about 
their structure.  We will show explicitly 
that supersymmetry is not spontaneously
broken in the limit we will consider; since instanton effects are turned
off in the limit of large $V$ and $\rho$, this  follows from
holomorphy of the superpotential and the decoupling of the pseudoscalar
partners of $V$ and $\rho$ at zero momentum, but  it is nice to have
an explicit check.  Also, in our computation,
we will derive the lower bound on $G_N$  stated above.  The computation
is a sort of strong coupling analog of a 
weak coupling, large radius expansion
of $(0,2)$ models that was carried out in \ref\twowittens{
L. Witten and E. Witten, ``Large Radius Expansion Of Superstring
Compactifications,'' Nucl. Phys. {\bf B281} (1987) 109.}.

Our computation is relevant to other problems in which
it is important to consider turning on $G$ in a supersymmetric way.
Such a problem is $M$-theory compactification on $\T^5/\Z_2$,
where five-branes\foot{To avoid confusion, let me stress that five-branes
in this paper will always be the 
five-branes of eleven-dimensional $M$-theory.  Thus, for instance
in compactification on $M^{10}\times \S^1/\Z_2$, the five-brane
propagates in bulk on the eleven-manifold, and is distinct from
$E_8\times E_8$ instantons, also often called five-branes, which
occur on the boundaries.  The two kinds of five-brane have some
couplings in common, but are definitely different as one has a tensor
multiplet on the world volume and one does not.}
 -- which are magnetic sources of $G$ -- appear
in the vacuum \ref\ugwitten{E. Witten, ``Five-Branes and $M$-Theory On
An Orbifold,'' hep-th/9512219.}; this compactification has also been studied
by Dasgupta and Mukhi  \ref\mukhi{K. Dasgupta and S. Mukhi, ``Orbifolds Of
$M$-Theory,'' hepth/9512196.}.  For this and for other
supersymmetric compactifications to six dimensions, we will obtain more 
precise
results than we get for compactification to four dimensions 
on Calabi-Yau threefolds; in compactification to six dimensions
the nonlinear structure is much simpler.
Among other things, we will get  a verification that the five-brane
configuration considered in \ugwitten\ in compactification
on $\T^5/\Z_2$ is compatible with supersymmetry, adding
support to the proposal made in that paper for the strong coupling
behavior of Type IIB superstrings on K3.  We will also describe
how to incorporate the five-branes in Calabi-Yau compactification.

While the strong coupling region we will study has the potential
to ameliorate  the usual puzzle about the low energy couplings,
we will not obtain any clear insight about the perhaps related puzzle,
stressed in \ref\olddine{M. Dine and N. Seiberg, ``Is The Superstring
Weakly Coupled?'' Phys. Lett. {\bf 162B} (1985) 299, ``Is The Superstring
Semiclassical?'' in {\it Unified String Theories}, eds. M. B. Green and
D. J. Gross (World Scientific, 1986).}
\nref\newdine{M. Dine and Y. Shirman, ``Truly 
Strong Couplings And Large Radius In String Theory'' 
to appear.} and revisited in \refs{\banks,\newdine},
of why supersymmetry breaking in a region where physics is computable
does not result in a runaway to weak coupling.   Thus, 
we will not obtain any particular insight about why there would be
a stable vacuum with broken supersymmetry (and no runaway to zero
$\alpha_{GUT}$) in a regime that can be approximated
by the description worked out below.     
Perhaps the occurrence of 
strong coupling in the second $E_8$ when Newton's
constant reaches its lower bound is a clue.

\newsec{Long Wavelength Expansion In Eleven Dimensions}

\def\M{{M^{11}}}
Our conventions in eleven dimensions will be those of \ref\berg{E. Bergshoeff,
M. de Roo, B. De Wit, and P. van Nieuwenhuysen, ``Ten-Dimensional
Maxwell-Einstein Supergravity, Its Currents, And The Issue Of Its
Auxiliary Fields,'' Nucl. Phys. {\bf B195} (1982) 97.}.  For example,
the signature of the space-time manifold $\M$
is $-+++\dots +$; gamma matrices obey $\{\Gamma_I,\Gamma_J\}
=2g_{IJ}$ and $\Gamma_1\Gamma_2\dots \Gamma_{11}=1$.  One
also defines $\Gamma_{I_1I_2\dots I_n}=(1/n!)\left(\Gamma_{I_1}\Gamma_{I_2}
\dots \Gamma_{I_n}\pm {\rm permutations}\right)$.  

The bosonic fields in eleven-dimensional supergravity are the metric
$g$ and a three-form $A$; the fermions are the spin $3/2$ gravitino
$\psi_{I\alpha}$, $\alpha$ being a spinor index.  
The field strength of $A$ is $G_{IJKL}=\partial_IA_{JKL}\pm
\,{23\,\,{\rm more}\,\,{\rm terms}}$.  
The supersymmetry transformation law for $\psi$ is
\eqn\purgo{\delta\psi_I=D_I\eta+ {\sqrt 2\over 288} (\Gamma_{IJKLM}
-8g_{IJ}\Gamma_{KLM})G^{JKLM}\eta.}
The condition for a spinor field $\eta$ to generate an unbroken
supersymmetry is that the right hand side of \purgo\ vanishes,
\eqn\consup{D_I\eta+ {\sqrt 2\over 288} (\Gamma_{IJKLM}
-8g_{IJ}\Gamma_{KLM})G^{JKLM}\eta=0.}
The most obvious way to obey this is to set $G=0$ and pick on $\M$ a metric
that admits a covariantly constant spinor field, $D_I\eta=0$.  
For example, in compactifying
from eleven to four dimensions on $X\times \S^1$, $X$ being a Calabi-Yau
manifold, one can take $G=0$ and then the covariantly constant spinors
on $X$ give unbroken supersymmetries.

Compactification on $X\times \S^1/\Z_2$ is more subtle because one
may not take $G=0$.  The reason for this  is that there is
a magnetic source of the $G$ field at the fixed points in $\S^1/\Z_2$.
In fact, while the Bianchi identity for $G$ on a smooth eleven-manifold
without five-branes or other singularities states 
that $dG=0$ (where $(dG)_{IJKLM}=\partial_IG_{JKLM}\pm {\rm cyclic
~permutations}$), there are several types of ``impurity''
that give contributions to $dG$.  For instance, $dG$ receives a delta
function contribution at the location of five-branes, and also
\ugwitten\ at certain codimension five singularities.  More directly
relevant for us is that in compactification on $\S^1/\Z_2$, there is
a delta function contribution 
to $dG$, supported at the $\Z_2$ fixed points.
If we consider the  fixed point set to be at $x^{11}=0$, then
the non-vanishing part of the fixed point contribution to $dG$
 is
\eqn\roaring{(dG)_{11\,IJKL}=
-{3\sqrt 2\delta(x^{11})\over 2\pi}\left(\kappa\over 4\pi\right)^{2/3}
\left(\tr F_{[IJ}F_{KL]}-{1\over 2}\tr R_{[IJ}R_{KL]}\right).}
Only one of the two $E_8$'s appears in \roaring, namely the one propagating
at $x^{11}=0$; that is the reason for the $1/2$ multiplying
$\tr R\wedge R$.  A formula such as the above
 is needed to reproduce in eleven dimensions the effects of
the string theory relation
$dH=-\sum_i\tr F_i\wedge F_i + \tr R\wedge R$; 
the precise coefficient in \roaring\
will be obtained elsewhere \newhorava.
In \roaring,   $\tr F\wedge F$ is as usual $1/30$ times the  trace
in the adjoint representation of $E_8$, and  
$\tr R\wedge R$ is the trace in the vector representation of $SO(1,10)$;
also,  $\tr F_{[IJ}F_{KL]}=(1/24)\tr F_{IJ}F_{KL}\pm {\rm permutations}$.
It is apparently impossible in Calabi-Yau compactification to find a 
vacuum with the property that $\tr \, F\wedge F -(1/2)\tr R\wedge R=0$
pointwise 
(the standard embedding of the spin connection in the gauge group
is a natural way to get $\tr\, F\wedge F-\tr R\wedge R=0$), so \roaring\
implies generically and perhaps always that $G\not= 0$ in Calabi-Yau
compactification.

Note that in studying physics on an orbifold, one can
either work ``upstairs,'' in this case on $M^{11}=\R^4\times X\times \S^1$
with $X$ a Calabi-Yau manifold,
and require invariance under the orbifolding group, in this case a $\Z_2$
that acts only on $\S^1$, or ``downstairs,'' directly on the quotient, in this
case $M^{11}/\Z_2=\R^4\times X\times \S^1/\Z_2$.  \roaring\ has been
written in the ``upstairs'' version, $G$ being a four-form on $M^{11}$ that
is odd under the $\Z_2$ ($G$ is odd because it changes sign under
orientation reversal);  in the analogous ``downstairs'' version,
explained in \newhorava, \roaring\ is replaced by a boundary condition
that has the same effect and in particular forces $G\not=0$.

Our goal, then, is to show in the first non-trivial order how $G$ can be
turned on to obey \roaring\ and preserve supersymmetry.
To be more specific, we work on $M^{11}/\Z_2=\R^4\times X\times \S^1/\Z_2$.  
According to \roaring, one can take $G$ to be
of order $\kappa^{2/3}$, so to lowest order in $\kappa$ (or in other words
to lowest order in an expansion in the inverse of the length scale of
$X\times \S^1/\Z_2$), we can set $G=0$.  For the starting point, then,
we take the metric on $M^{11}/\Z_2$ to be the product of a flat metric
on $\R^4\times \S^1/\Z_2$ with a Calabi-Yau metric on $X$.  The unbroken
supersymmetries come from covariantly constant spinor fields on $X$.
Then in order $\kappa^{2/3}$, we must pick holomorphic $E_8$ bundles
on the $\Z_2$ fixed points (note from \mippy\ that the gauge kinetic energy
is of order $\kappa^{2/3}$ compared to the gravitational kinetic energy,
so that as in the study of Type I superstrings the choice of a gauge bundle
can be considered a kind of higher order correction relative to the
basic choice of gravitational vacuum).
Also in order $\kappa^{2/3}$, we find a solution of the equations of motion
for $G$, including the source term \roaring. 
Then modifying also the metric
on $M^{11}$ and the spinor field $\eta$, we  aim to  
obey the condition \consup\ of unbroken supersymmetry in order $\kappa^{2/3}$.

The first step is then to find a solution of the equations of motion for
$G$ together with the Bianchi identity.  The equation of motion contains
a term involving $G\wedge G$, but this will vanish in the situations
we will consider, since (given the four-dimensional Poincar\'e invariance
and the vanishing of $G_{1234}$), there is no ``room'' for a non-zero
eight-form $G\wedge G$.  The equation of motion for $G$ then reduces
simply to
\eqn\hurry{D^IG_{IJKL}=0,}
and the  Bianchi identity reads
\eqn\flurry{D_IG_{JKLM}\pm {\rm permutations}={\rm sources},}
where the ``sources'' are delta functions supported at $\Z_2$ fixed points
or at five-branes.  The condition that the Bianchi identity has a solution
is that the source terms add up to zero cohomologically.  If no five-branes
are present, this condition simply amounts to the standard string theory
constraint on the   total $E_8\times E_8$ instanton number
(thus, $\sum_{i=1,2}\tr F_i\wedge F_i-\tr R\wedge R$ must vanish 
cohomologically,
as in perturbative string theory where this expression equals $-dH$).
\hurry\ and \flurry\ then have a solution which becomes unique if one asks
that $G$ comes from a four-form on $X\times \S^1$ that is odd under the $\Z_2$
action on $\S^1$ and cohomologically trivial.\foot{A cohomologically 
non-trivial 
and $\Z_2$ odd addition to $G$ would be of the form
$H\wedge dx^{11}$, with $H$ a harmonic form on $X$.  This addition breaks
supersymmetry, just as turning on a cohomologically non-trivial three-form
field $H$ in the weakly coupled heterotic string breaks supersymmetry.}
With this $G$ as a starting point, we will verify that the conditions
for unbroken supersymmetry have a unique solution up to order $\kappa^{2/3}$.

 \nref\duff{M. Duff, R. Minasian, and E. Witten, 
``Evidence For Heterotic/Heterotic
Duality,'' hep-th/9601036.}
It is interesting to consider the general case in which the source terms
in the Bianchi identity come from five-branes as well as $\Z_2$ fixed points.
This generalization involves issues that are less familiar from the string
theory viewpoint.  To preserve supersymmetry (roughly as in
\ref\strom{K. Becker, M. Becker, and A. Strominger,
``Five-Branes, Membranes, And Nonperturbative String Theory,''
Nucl. Phys. {\bf B456} (1995) 130.}), take the $\alpha^{th}$  
five-brane (whose world-volume  should be a submanifold of $M^{11}$ of 
codimension
five) to be located at the product of a holomorphic curve $C_\alpha\subset X$ 
times a point $P_\alpha\in \S^1/\Z_2$.  One must then include the  five-brane 
source
terms in the Bianchi identity for $G$.  The condition for the Bianchi
identity to have a solution is now that (cohomologically)
\eqn\anomcan{{\sum_i\tr F_i\wedge F_i-\tr R\wedge R\over 8\pi^2} 
+\sum_\alpha[C_\alpha]=0.}
with $[C_\alpha]$ the Poincar\'e dual cohomology class to $C_\alpha$. 
(The fact that the coefficient of $[C_\alpha]$ in this formula is precisely
$+1$ follows from the observation in \duff\ that the five-brane  
and Yang-Mills instanton make the same contribution to the irreducible
part of the gravitational anomaly.)

The reason that there is no difficulty in including fivebranes in our
computation is really the following.  
Let us decompose the space of complex-valued $p$ forms on $X\times \S^1$
into forms
of type $(a,b,c)$, where $a+b+c=p$, and $a$ counts the number of holomorphic
indices tangent to $X$, $b$ counts the number of antiholomorphic indices
tangent to $X$, and $c$ is 1 or 0 depending on whether a factor
$dx^{11}$ is present or absent.  Then in our computation, the
only important fact about \roaring\ will be  that the source term for $dG$  is
supported at singularities and 
is a $(2,2,1)$ form.  Inclusion of five-branes causes no trouble if
the five-brane source for $dG$ is likewise of type $(2,2,1)$,
which we ensure by locating  the $\alpha^{th}$ five-brane
 at $C_\alpha\times P_\alpha$ with
$C_\alpha$ and $P_\alpha$ as above.  Given this restriction, we need not
mention whether five-branes are present or not (except in  quantitatively
estimating the lower bound on $G_N$).

\nref\liu{M. J. Duff, R. Minasian, and J. T. Liu, ``Eleven-Dimensional
Origin Of  String/String Duality,'' Nucl. Phys. {\bf B452} (1995) 261.}
Incidentally, this formalism makes it manifest that (as partly
verified in \duff) the vacua with five-branes
present are free of space-time anomalies, as long as \anomcan\ holds.
   Since
there are no gravitational anomalies on a smooth eleven-manifold,
the  anomalies
in eleven dimensions are localized at five-branes and singularities, but
the anomaly on the five-brane world-volume cancels \refs{\liu,
\ugwitten}, and the same is true at the $\Z_2$ orbifold singularities
\newhorava.  The space-time anomaly cancellation can thus be understood
locally in eleven dimensions, with no need to verify low energy details.

\subsec{The Computation}

In performing the computation on $M^{11}=\R^4\times X\times \S^1$,
we adopt the following conventions.  Indices 
$I,J,K,\dots$ from the middle of the alphabet
run from 1 to 11 and are tangent to $M^{11}$.  Indices $X,Y,Z,\dots$
from the end of the alphabet run from 5 to 11 and are tangent
to $X\times \S^1$.  Indices $A,B,C,\dots$ from the beginning of the
alphabet are tangent to $X$; 
we also  use $a,b,c,\dots =1\dots 3$ for holomorphic indices tangent to $X$,
and $\bar a, \bar b, \bar c,\dots =1 \dots 3$ for analogous antiholomorphic
indices. The index 11 is tangent to $\S^1$.  Meanwhile,
$\mu,\nu,\lambda,\,\dots = 1\dots 4$ will be  tangent to $\R^4$.

We take the metric on $\R^4$ to be the Minkowski metric $\eta_{\mu\nu}dx^\mu
dx^\nu$ and the metric
on $\S^1$ to be $(dx^{11})^2$. (The $\S^1$ has circumference $2\pi\rho$.)
The Calabi-Yau manifold $X$ has a metric tensor $g_{AB}$ with
non-zero components $g_{a\bar b}=g_{\bar b a}$ and
a Kahler form $\omega_{AB}$ with non-zero components
$\omega_{a\bar b}=-i g_{a\bar b}=-\omega_{\bar b a}$.  The
gamma matrices tangent to $X$ are $\Gamma^a$ and $\Gamma^{\bar b}$ with
$\{\Gamma^a,\Gamma^{\bar b}\}=2g^{a\bar b}$ and other anticommutators
vanishing.  We will look for a solution of the supersymmetry condition
\consup\ that obeys $\Gamma^{11}\eta=\eta$ (so as to be invariant
under the $\Z_2$ projection), and $\Gamma^a\eta=0$; the complex
conjugate of $\eta$ would obey $\Gamma^{11}\eta'=\eta'$ 
and $\Gamma^{\bar b}\eta'=0$.

We begin, as explained above, with the cohomologically trivial solution
$G_{IJKL}$ of the Bianchi identity $dG={\rm ``sources}$''
and equations of motion $D^IG_{IJKL}=0$.
Because the source term in the Bianchi identity is of type $(2,2,1)$,
the non-zero components of $G$ are $G_{ABCD}$, of type $(2,2,0)$,
and $G_{ABC\,11}$, a mixture of forms of types $(2,1,1)$ and $(1,2,1)$.
\foot{A $(3,0,1)$ part of $G$ would be of the form $H\wedge dx^{11}$,
with $H$ a $(3,0)$ form on $X$.  Given that the sources are of type $(2,2,1)$,
the Bianchi identity $dG={\rm sources}$ implies that the $(3,1,1)$ part
of $dG$ vanishes so that $\bar\partial H=0$.  The equation of motion
similarly implies that $H$ is independent of $x^{11}$.
Since $G$ is cohomologically trivial, the $x^{11}$-independent holomorphic
three-form $H$ must vanish.
The $(0,3,1)$ part of $G$ vanishes likewise.}
It is convenient to introduce
\eqn\conven{\eqalign{\beta_A & = \omega^{BC}G_{ABC\,11}\cr
                     \theta_{AB} & = G_{ABCD}\omega^{CD}\cr
                     \alpha & = \omega^{AB}\theta_{AB}=
                     \omega^{AB}\omega^{CD}G_{ABCD}. \cr}}
These objects obey various identities that follow from the equations
of motion and Bianchi identities of $G$.  We will write these equations
as they hold away from the singularities and five-branes, so we can
ignore the source terms in the Bianchi identities.  
The $(1,3,1)$ part of the Bianchi identity gives
\eqn\rumbo{D_{\bar a} G_{a\bar b\bar c \,11}+D_{\bar b}G_{a\bar c\bar a\,11}
+D_{\bar c}G_{a\bar a\bar b\,11}=0.}
The equation of motion $ D^IG_{I\bar a\bar b\, 11}=0$, when combined
with the vanishing of the $(0,3,1)$ part of $G$, gives
\eqn\humm{0=\omega^{a\bar a}D_{\bar a}G_{a\bar b\bar c \,11}=0.}
If one contracts \rumbo\ with $\omega^{a\bar a}$ and uses \humm, one
geta
\eqn\humbo{D_{\bar a}\beta_{\bar b}-D_{\bar b}\beta_{\bar a} = 0.}
The $(2,3,0)$ part of
the Bianchi identity, that is the condition $\partial_{\bar a}G_{b\bar bc\bar 
c}
+\partial_{\bar b}G_{b\bar c c\bar a}+\partial_{\bar c}G_{b\bar a c\bar b}=0$,
can if contracted with $g^{c\bar c}$, be reduced after using the equations
of motion to 
\eqn\redto{0=-{i\over 2} \left(\partial_{\bar a}\theta_{b\bar b} -\partial_{
\bar b}
\theta_{b\bar a}\right)+D^{11}G_{b\bar a\bar b \,11}.}
If this is contracted with $\omega^{b\bar b}$ one gets
\eqn\edto{0=-{i\over 2}D^A\theta_{A\bar b}+{1\over 4} D_{\bar b}\alpha
-{i\over 2} D_{11}\beta_{\bar b}.}
One component of the equation of motion is $D^{11}G_{a\bar a\bar b \,11}+D^b
G_{a\bar a\bar b b}=0$; if this is contracted with $\omega^{a\bar a}$ one gets
\eqn\uredto{0=D_{11}\beta_{\bar b}-D^b\theta_{b\bar b}.}
Combining \edto\ and \uredto, we obtain
\eqn\nedto{D_{11}\beta_{\bar a}=-{i\over 4} \partial_{\bar a}\alpha.}
Also, upon starting with the equation of motion $D^AG_{ABC\,11}=0$
and contracting with $\omega^{BC}$, one learns
\eqn\medto{D^A\beta_A=0.}  Finally, starting with the component
$dG_{ABCD\,11}=0$ of the Bianchi identity and
contracting with $\omega^{AB}\omega^{CD}$, one learns that 
$\partial_{11}\alpha$ is a total derivative with respect to $x^A$,
as a result of which
\eqn\udto{ 0={\partial\over\partial x^{11}} \int_X\alpha \sqrt g d^6x.}

To evaluate the $\Gamma G \eta$ terms in the supersymmetry condition
\consup, one needs the following identities:
\eqn\kedto{\eqalign{\Gamma^{XYZW}G_{XYZW}\eta & 
=\left(-3\alpha-12i\beta_{\bar b}
\Gamma^{\bar b}\right) \,\eta \cr
\Gamma^{XYZ}G_{11\,XYZ}\,\eta & = 3i\Gamma^{\bar b}\beta_{\bar b}\eta 
\cr
\Gamma^{XYZ}G_{a XYZ} & = \left(-3i\beta_a+3G_{a\bar b\bar c\,11}\Gamma^{\bar 
b\bar
 c}-3i\theta_{a\bar b}\Gamma^{\bar b} \right)\eta \cr
\Gamma^{XYZ}G_{\bar b XYZ} & = -3i\bar \beta_{\bar b}\eta.\cr}}
These identities use $\Gamma^{11}\eta=\eta$, $\Gamma^{b}\eta=0$.
Using these identities, one can compute that the $\Gamma G\eta$ terms
in \consup\ are 
\eqn\hedto{\eqalign{
{\sqrt 2\,dx^I\over 288 }&\left(\Gamma_{IJKLM}-8g_{IJ}\Gamma_{KLM}\right)
G^{JKLM}\eta= {\sqrt 2\over 288} 
\left(dx^{11}\left(-3\alpha -24i\beta_{\bar b}\Gamma^{\bar b}\right)
+dx^{\bar b} \cdot 12 i 
\beta_{\bar b}\right.\cr &\left. +dx^a\left(36 i \beta_a
+\left(36 i \theta_{a\bar b}-3\alpha g_{a\bar b}\right)\Gamma^{\bar b}
+\left(-36G_{a\bar a\bar b\,11}-6i(g_{a\bar a}\beta_{\bar b}-g_{a\bar b}
      \beta_{\bar a}\right) \Gamma^{\bar a\bar b}\right)\right.\cr &\left. 
+dx^\mu\Gamma_\mu\left(-3\alpha-12i\beta_{\bar b}\Gamma^{\bar b}\right)
\right)\eta.\cr}}
 
To evaluate \consup, we also need to look at terms that come from
a change in the metric on $M^{11}$.  A careful study of the equations
shows that one can take the perturbation in the metric to be block
diagonal and to obey certain additional restrictions, 
so that the perturbed line element is
\eqn\rully{ds^2=\eta_{\mu\nu}dx^\mu dx^\nu(1+b)+2(g_{a\bar b}+h_{a\bar b})
dx^adx^{\bar b}+(1+\gamma)(dx^{11})^2+\dots ,}
where $b$, $h$, and $\gamma$ will be of order $G$, that is of order
$\kappa^{2/3}$, and the $\dots$ are terms of order $\kappa^{4/3}$.
To preserve four-dimensional Lorentz invariance, $b,h$, and $\gamma$
are functions only of the $x^Y$, $Y\geq 5$.  
Note, for example, that the metric on $X$ is still
hermitian, but will no longer be Kahler as we will see.  
We also replace the spinor field $\eta$ with 
\eqn\pillow{\tilde\eta= e^{-\psi}\eta,}
$\psi$ being again of order $\kappa^{2/3}$.

In general, under a perturbation
of the metric $g_{IJ}\to g_{IJ}+h_{IJ}$, to first order in $h$, the
covariant derivative of a spinor changes by $D_I\eta
\to D_I\eta-{1\over 8} (D_Jh_{KI}-D_Kh_{JI})\Gamma^{JK}\eta$.
Using this formula, one finds that, with $\eta$ covariantly
constant with respect to the unperturbed metric, one has to first order
in the perturbation
\eqn\truly{\eqalign{dx^ID_I\tilde \eta=& 
\left(-dx^Y\partial_Y\psi +
dx^\mu\Gamma_\mu\left( {1\over 4}\partial_{11}b
+{1\over 4} \partial_{\bar a}b\Gamma^{\bar a} \right) \right.
\cr &\left.  -dx^a\left(-{1\over 4}\partial_{11}h_{\bar a a}
      \Gamma^{\bar a} +{1\over 8}\left(\partial_{\bar a}h_{\bar b a}
     -\partial_{\bar b}h_{\bar a a} \right)\Gamma^{\bar a\bar b}
+{1\over 4}g^{b\bar b}\partial_bh_{a\bar b}\right)\right.\cr
& \left. -dx^{\bar a}\left(
-{1\over 4}g^{a\bar b}\partial_{\bar b}h_{a\bar a}\right)
-{dx^{11}\over 4}\partial_{\bar a}\gamma \Gamma^{\bar a}\right)\tilde
\eta .\cr}}

The condition
\eqn\rocon{D_I\tilde \eta+ {\sqrt 2\over 288} (\Gamma_{IJKLM}
-8g_{IJ}\Gamma_{KLM})G^{JKLM}\tilde\eta=0}
can now be evaluated.  
The terms proportional to $\Gamma^{\bar a\bar b}\eta$ give
\eqn\jumpy{\partial_{\bar a}h_{b\bar b}-\partial_{\bar b}h_{b\bar a}
=-\sqrt 2\left(G_{b\bar a\bar b \,11}+{i\over 6}\left(g_{b\bar a}\beta_{\bar b}
-g_{b\bar b}\beta_{\bar a}\right) \right) .}
The terms proportional to $\Gamma^{\bar a}\eta$ give two equations:
\eqn\funeq{\eqalign{\beta_{\bar a} & = 
       {3i\over \sqrt 2} \partial_{\bar a}\gamma   \cr
        \partial_{11}h_{a\bar b} & =-{1\over \sqrt 2}\left(i\theta_{a\bar b}
-{1\over 12} \alpha g_{a\bar b}\right).\cr}}
The terms involving $\eta$ multiplied by a function, without gamma
matrices, give
\eqn\longive{\eqalign{0=&
-dx^Y\partial_Y\psi+dx^a\left(-{1\over 4}g^{b\bar b}\partial_b
h_{a\bar b} \right) + dx^{\bar a} \left({1\over 4} g^{a\bar b}\partial_{\bar b}
h_{a\bar a}\right)\cr & +{\sqrt 2\over 288}\left( dx^{11}(-3\alpha) 
+dx^{\bar b}\cdot 12i\beta_{\bar b}+dx^a\cdot 36i\beta_a\right).\cr}}
The imaginary part of this equation gives
\eqn\perfume{ g^{b\bar b} \partial_bh_{a\bar b} ={\sqrt 2 i\over 3} \beta_a.}
The real part gives
\eqn\regives{\eqalign{ \partial_{11}\psi+{\sqrt 2\over 96}\alpha &= 0 \cr
              \partial_{\bar a}\psi+{\sqrt 2 i\over 24}\beta_{\bar a} &=0.\cr}}
which determine $\psi$ in terms of $\alpha$ and $\beta$, that is in terms of
$G$, up to an irrelevant additive constant.  
However, we need to verify that (with a suitable choice of the constant)
the solution $\psi$ of those equations is real.  To prove that ${\rm Im}\,\psi$
is a constant, it suffices to prove that $g^{a\bar b} \partial_a\partial_{\bar 
b} 
{\rm Im}\,\psi=0$.  In fact, using the second equation in \regives\ (and
its complex conjugate) together with \medto,
one gets
\eqn\plly{g^{a\bar b}\partial_a\partial_{\bar b}{\rm Im}\,\psi=
g^{a\bar b}\partial_{\bar b}\partial_a\left({\psi-\bar\psi\over 2i}\right)
={1\over 24\sqrt 2}D^A\beta_A=0.}
Furthermore,
comparison of the first equation in \funeq\ to the second in \regives\
shows that
\eqn\hully{ \gamma= 8\psi +f(x^{11}).}
It is impossible to determine $f$, as $f$ can be changed arbitrarily
by a reparametrization of $x^{11}$, which has not yet been fixed.
One natural coordinate condition would be $f=0$; another would
be
\eqn\bully{\int_X \gamma \sqrt g d^6x = 0.}
Finally, the terms proportional to $\Gamma_\mu$ give
\eqn\joeqns{\eqalign{\beta_{\bar a} &= - {6i\over \sqrt 2}\partial_{\bar a} b
\cr
\partial_{11} b & = {\sqrt 2\over 24}\alpha.}}
Comparing to the above this implies
\eqn\rumble{b=-4\psi}
up to an irrelevant additive constant.

What remains is to determine $h$.  The three equations that we have not
yet used in determining $\psi$, $\gamma$, and $b$ may be collected
as follows
\eqn\threemore{\eqalign{\partial_{\bar a}h_{b\bar b}-\partial_{\bar b}h_{b\bar 
a}
& =-\sqrt 2\left(G_{b\bar a\bar b \,11}+{i\over 6}\left(g_{b\bar a}\beta_{\bar 
b}
-g_{b\bar b}\beta_{\bar a}\right) \right)\cr
g^{\bar b b}\partial_{\bar b}h_{b\bar a} & 
                         =-{\sqrt 2 i\over 3}\beta_{\bar a} \cr
\partial_{11}h_{a\bar b} & =-{1\over \sqrt 2} \left(i\theta_{a\bar b} -{1\over 
12}
\alpha g_{a\bar b} \right)\cr}}
and must serve to determine $h$.  The integrability condition
for the first equation in \threemore\ is that the right hand
side should be annihilated by $\bar \partial$.  This is true according
to \rumbo\ and \humbo.  Given this,
according to standard Hodge theory, at fixed $x^{11}$,
the first two equations in \threemore\
have a common solution which is unique up to the possibility of
adding to $h$ a harmonic $(1,1)$ form on $X$.  The $x^{11}$ dependence
is then to be determined from the last equation in \threemore.  This
will determine $h$ uniquely up to the possibility of adding an
$x^{11}$-independent harmonic $(1,1)$ form, which is the expected
ambiguity corresponding to a displacement of the Kahler moduli of the vacuum.
The only remaining point is that the last equation in \threemore\
is compatible with the first two.  Compatibility of the first
and third equations in \threemore\ is the statement that $\partial_{11}$
of the right hand side of the first equation equals $\bar \partial$
of the right hand side of the third.  This can be verified using \redto, 
\uredto, and \nedto.  Compatibility of the second and third equations in
\threemore\ can likewise be verified using 
\uredto\ and \nedto.

\subsec{Lower Bound On Newton's Constant}

Now, as anticipated in the introduction,
let us try to estimate the lower bound on Newton's constant subject
to the validity of the calculation that we have performed.
We let $v(x^{11})=\int_X\sqrt g d^6x$ be the volume of $X$ at given
$x^{11}$.  As explained in the introduction, the GUT
coupling is $\alpha_{GUT}=(4\pi\kappa^2)^{2/3}/2V$, where $V$ is
the value of $v$ at the $\Z_2$ fixed point.  However, $v$ will have different
values at the two fixed points $x^{11}=0$ and $x^{11}=\pi \rho$.
We set $V=v(0)$ and undertake to compute $v(\pi\rho)$.

The starting point is simply that
\eqn\jodo{{\partial\over\partial x^{11}}v=\int_X {1\over 2}g^{AB}
\partial_{11}h_{AB} \sqrt g d^6x=\int_X g^{a\bar b}\partial_{11} h_{a\bar b}
\sqrt g d^6x.}
So, using the last equation in \threemore, 
\eqn\minorly{{\partial\over\partial x^{11}} v =-{1\over 4\sqrt 2}\int_X
\alpha \sqrt g d^6x.}
According to \udto, the right hand side is independent of $x^{11}$, 
so $v$ will vary linearly in $x^{11}$.  
  Generically, $\alpha$ will be non-zero.
If, for example, $\alpha$ is positive, then $v(0) >v(\pi\rho)$.
The $E_8$ which is located at $x^{11}=0$ is therefore the more weakly
coupled of the two.  If we understand $\alpha_{GUT}$ to be the 
coupling of the more weakly coupled of the two $E_8$'s, then 
if one keeps $\alpha_{GUT}$ fixed and increases $\rho$, the coupling
of the second $E_8$ will diverge at a finite value of $\rho$, where $v=0$.
Newton's constant, meanwhile, decreases with increasing $\rho$.
The smallest possible value of Newton's  constant, in a regime
in which the eleven-dimensional calculation is valid, is the  
value at which $v=0$ and the coupling in the second $E_8$ diverges.
The occurrence at this point of ``infinite bare coupling'' in one $E_8$
is extremely interesting; it is reminiscent
of an unexpected failure of perturbation theory found in \polwitten,
and also is related to the phase transition at finite heterotic string
coupling noted in \duff\ (which arises in just the same way in ${\rm K3}
\times \S^1/\Z_2$ compactification, as will be clear in the next
section).

Since $\alpha$ is independent of $x^{11}$, we can determine it by
looking at the limit as $x^{11}$ approaches zero from above.  According
to \newhorava, the limiting value of $G_{ABCD}$ is
\eqn\gurtle{G_{ABCD}=-{3\over 2\pi\sqrt 2}\left({\kappa\over 4\pi}\right)^{2/3}
\left(     F_{[AB}F_{CD]}-{1\over 2}R_{[AB}R_{CD]}\right).}
This implies that
\eqn\hurtle{\sqrt g \alpha = \sqrt g \omega^{AB}\omega^{CD}G_{ABCD}
=-{2\over \pi \sqrt 2} \left({\kappa\over 4\pi}\right)^{2/3} \omega
\wedge \left(\tr F\wedge F-{1\over 2}\tr R\wedge R\right).}
So 
\eqn\purple{{\partial v\over \partial x^{11}}=2\pi\left({\kappa\over 4\pi}
\right)^{2/3} \int_X\omega\wedge {\tr F\wedge F-{1\over 2}\tr R\wedge R
\over 8\pi^2}}
and hence in this approximation
\eqn\nurble{V'=v(\pi \rho)=V+2\pi^2\rho\left({\kappa\over 4\pi}\right)^{2/3}
\int_X\omega\wedge { \tr F\wedge F-{1\over 2}\tr R\wedge R
\over 8\pi^2}.}
Note that, if the ``instanton number'' is larger at $x^{11}=0$ than
at $\pi\rho$, the integral on the right hand side of \nurble\ is
negative; in fact it follows from the supersymmetric relation
$\omega^{AB}F_{AB}=0$ that $\omega\wedge \tr F\wedge F$ is negative.
The critical value of $\rho$, at which $V'=0$, is thus in this approximation
\eqn\purple{\rho_c={V\over 2\pi^2\left({\kappa\over 4\pi}\right)^{2/3}
      \left|\int_X \omega\wedge{\tr F\wedge F-{1\over 2}\tr R\wedge R
\over 8\pi^2}\right|}.}

Let $W$ be the volume of $X\times\S^1$.  Newton's constant is
\eqn\kolp{G_N=\kappa^2/8\pi W} (this formula is given in \kippy, using
the tree level expression $W=2\pi\rho V$).  In the approximation of
taking $v$ to vary linearly (strictly valid only to lowest non-trivial
order in $\kappa^{2/3}$), the volume at the critical point is 
$W_c=\pi V\rho_c$.  (This comes by multiplying the circumference,
$2\pi\rho$, of the $\S^1$ times the average value $V/2$ of the K3 volume.)
This upper bound on $W$, when inserted in \kolp, gives the lower bound
on $G_N$ claimed in \ippo.

Since we treated the supergravity only in a linearized approximation,
the precise critical values we estimated for $\rho$, $W$, and $G_N$
are uncertain to within factors of order one.  However, it is
possible, by looking at the gauge couplings, to be more precise
about the fact that there is a breakdown of the low energy
supergravity, at roughly the point found above.  The inverse
of the $E_8$ gauge coupling at $x^{11}=0$ is, from  \kippy,
\eqn\jumbo{{1\over \alpha_{GUT}} = {2V\over (4\pi\kappa^2)^{2/3}}.}
The inverse of the second $E_8$ gauge coupling $\tilde\alpha$
is given by a similar formula with $V$ replaced by $V'$, so from
\nurble\
\eqn\numbo{{1\over \tilde\alpha}={2(V-\rho\sum_ac_a\omega_a)\over
(4\pi\kappa^2)^{2/3}},}
with $\omega_a$ the periods of $\omega$ and $c_a$ certain constants
that can be found from \nurble.  The number of independent
periods of $\omega$ is $h=\dim H^{1,1}(X)$.  The pseudoscalar
partners of $V$ and   the $\omega_a$ are ``axions'' that decouple
at zero momentum in the approximation of eleven-dimensional
supergravity (where one ignores Yang-Mills instantons and also
the membrane instantons that are related to world-sheet instantons
of string theory).  From $V$ and the $\omega_a$ one can make
$h+1$ functions $r_\lambda$, $\lambda=0,\dots, h$,
which (up to linear transformations and addition of constants) 
are determined
uniquely by the following conditions: they are    real parts of
chiral superfields and are invariant under shifts of the axions.
The usual  constraints of holomorphy together with the axion
 decoupling  imply that $1/\tilde\alpha$ must be a linear combination
of these functions.  
From \numbo, we see that in the approximation
that we have considered, the $r_\lambda$ are $r_0=V$ and
$r_a= \rho\omega_a$, $a=1,\dots, h$.  
\numbo\ suffices to determine the coefficients
when $1/\tilde\alpha$ is expressed as a linear combination
of the $r_\lambda$, and shows that in the supergravity approximation
$1/\tilde\alpha$ goes to zero at a certain finite value of the 
$r_\lambda$.  At that point, the low energy supergravity
approximation breaks down, and instantons in the second $E_8$ must
be taken into account.

\newsec{Compactification To Six Dimensions}
\def\S{{\bf S}}

In this section, we will apply similar ideas to supersymmetric
compactifications from eleven to six dimensions with non-vanishing
$G$ field.  The main examples are compactification on
${\rm K3}\times \S^1/\Z_2$, related to the heterotic string on K3,
compactification on $\T^5/\Z_2$, related 
\nref\dasgupta{K. Dasgupta and S. Mukhi, ``Orbifolds Of $M$ Theory,''
hep-th/9512196.}
\nref\otherwitten{E. Witten, ``Five-branes And $M$ Theory On An
Orbifold,'' hep-th/9512219.}
\refs{\dasgupta,\otherwitten} to Type IIB on K3, and various other
$\Z_2$ orbifolds of ${\rm K3}\times \S^1$ \ref\asen{A. Sen, ``$M$-Theory
On ${\rm K3}\times {\bf S}^1/{\bf Z}_2$,'' to appear.}, 
some apparently
related 
to K3 orientifolds discussed in \ref\atish{A. Dabholkar and J. Park,
``An Orientifold Of Type IIB Theory On K3,'' hepth/9602030.} and
$F$-theory compactifications \ref\vafanew{C. Vafa, ``Evidence For
$F$-Theory,'' hepth/9602022}.
On ${\rm K3}\times \S^1/\Z_2$, five-branes can be included if one
wishes \duff, and in the other examples five-branes must
be included \otherwitten\ to neutralize the overall magnetic charge.

There are three types of source for the $G$ field.  A five-brane is
a delta function source of strength 1, a codimension five $\Z_2$
orbifold singularity is a source of strength $-1/2$ \otherwitten,
and a codimension one $\Z_2$ orbifold singularity contributes a source
term, proportional to $\tr F\wedge F-(1/2)\tr R\wedge R$, which figured
in the last section.

Despite the diversity of examples, they can all be treated together.
Moreover, the results are much simpler than for compactification to
four dimensions, and it is straightforward to go beyond the linearized
approximation used in the last section and find the exact
supersymmetric solution of the supergravity theory to all orders in
$\kappa$.  This gives a description that is valid even when
the corrections from turning on $G$  are big, as long as all the relevant
length scales are large compared to the eleven-dimensional Planck
length.  That condition breaks down under certain conditions,
but treating the non-linear terms will enable us to get a better
understanding of how it breaks down and thus a better understanding
of how the strong coupling conundrum explained in \duff\ appears
in this framework.

\def\K{{\rm K}}
\def\K3{{\rm K3}}
\def\M{{M^{11}}}
We will work  
on $M^{11}=\R^6\times {\rm K}$, where K may be ${\rm K3}\times \S^1$
or $\T^5$.  We really want a $\Z_2$ orbifold of $\R^6\times {\rm K}$ (with
the $\Z_2$ acting on K only),
which we describe by giving a $\Z_2$-invariant configuration
on $\R^6\times {\rm K}$.  We write the starting line element on
$\M$ (valid to lowest order in $\kappa$, before turning on $G$) as
\eqn\loko{ds^2_{(0)}=\eta_{\mu\nu}dx^\mu dx^\nu+g_{AB}dx^Adx^B,}
where $x^\mu$, $\mu=1,\dots,6$ are the coordinates of $\R^6$ and
$x^A$ are local coordinates on K; the first term is the Minkowski
metric on $\R^6$, and the second
is  a standard metric on K (that is, the metric
on K is the product of a hyper-kahler metric on K3 times
a flat metric on $\S^1$ if ${\rm K}={\rm K3}\times \S^1$, or is a flat metric
on ${\rm K}$ if ${\rm K}=\T^5$).  
Indices $\mu,\nu$ will run from $1$ to 6, $A,B,C\dots$
from $5$ to 11, and $I,J,K,\dots$ from 1 to 11.
  \loko\ is the metric on $\M$ to lowest order in $\kappa$,
with the $G$ field ignored.  With $G$ turned on, it will
turn out that the metric is of the form
\eqn\looko{ds^2=e^b\eta_{\mu\nu}dx^\mu dx^\nu+e^fg_{AB}dx^Adx^B,}
where $b$ and $f$ are functions on K that will be determined.

As in the last section, the starting point is to determine $G_{ABCD}$
from the equations
\eqn\theeqns{\eqalign{ D^AG_{ABCD} & = 0 \cr
                        dG & = {\rm sources}.\cr}}
There is a subtlety here which is the key to eventually obtaining
a simple solution.  The first equation in \theeqns\ depends on
the metric \looko\ on $\M$, which we do not yet know; we only
know the starting metric \loko.  It turns out, however, that
if we determine $G_{ABCD}$ (as a differential form, that is
with all indices down!) using the starting
metric, then the {\it same} $G$ will solve \theeqns\ in the exact metric.
This depends on the following circumstance.  Solving \theeqns\ in the
starting metric is equivalent to finding a function $w$ such that
\eqn\yess{G_{ABCD}=-\epsilon^{(0)}_{ABCDE}\partial^Ew}
-- with $\epsilon^{(0)}$ the completely antisymmetric tensor
in the metric $ds^2_{(0)}$ -- 
such that 
\eqn\polly{\nabla_0 w = {\rm sources},}
where $\nabla_0$ is the Laplacian in the original metric, and the ``sources''
in \polly\ are derived from the sources in the Bianchi identity
for $G$, that is the second equation in \theeqns.  In particular, \polly\
determines $w$ uniquely up to an irrelevant additive constant.
Changing the metric on K will not affect the Bianchi identity,
which is completely metric-independent.  It will, however,
change the equations of motion.  In the corrected metric \looko,
the relation of $G$ to $w$ is
\eqn\besso{G_{ABCD}=-e^{-3f/2}\epsilon_{ABCDE}\partial^Ew}
where now $\epsilon$ is the $\epsilon$ tensor in the corrected metric.
The equation of motion for $G$ -- that is, the first equation in \theeqns\ --
then becomes
\eqn\yusso{d(e^{-3f/2}dw) = 0.}
For general $f$, this would not hold, but when we actually solve
for $f$, we will find that $f=F(w)$ for some function $F$, which
ensures that \yusso\ holds.  To recapitulate then, we first
solve \theeqns\ and determine $w$; then we find the corrected
metric, determining $b$ and $f$, which will be  
such that \theeqns\ holds also in the new metric.

For any of the compactifications of interest, the unbroken supersymmetries,
in the limit of ignoring $G$, are generated by covariantly constant
spinor fields $\eta$ that obey
\eqn\yulu{\Gamma_{ABCDE}\eta=\epsilon_{ABCDE}\eta.}
(For the case of compactification on $\K3\times \S^1/\Z_2$, one also
has $\Gamma_{11}\eta=\eta$, but we will not need to use this.)
As in section two, we first compute
\eqn\ruppo{\eqalign{
{\sqrt 2\over 288}dx^I\left(\Gamma_{IJKLM}-8g_{IJ}\Gamma_{KLM}
\right) G^{JKLM}\eta 
= {\sqrt 2e^{-3f/2}\over 288}&  \left(-24dx^A\partial_Aw-24
dx^\mu\Gamma_\mu \Gamma^A\partial_Aw \right.\cr &\left. 
+48dx^A\Gamma_{AB}\partial^Bw\right)\eta.\cr}}
Then, introducing a rescaled spinor field $\tilde \eta=e^{-\psi}\eta$,
we compute that the covariant derivative of $\tilde \eta$ in the
corrected metric \looko\ is
\eqn\dummy{dx^ID_I\tilde\eta =\left(-dx^A\partial_A\psi
-{1\over 8}dx^\mu\Gamma_\mu\Gamma^A\partial_Ab-{1\over 8}dx^A\partial^Bf
\Gamma_{AB}\right)\tilde\eta.}
The condition \consup\ of unbroken supersymmetry then gives
\eqn\ruffy{\eqalign{-dx^A\partial_A\psi
-&{1\over 8}dx^\mu\Gamma_\mu\Gamma^A\partial_Ab-{1\over 8}dx^A\partial^Bf
\Gamma_{AB}\cr &+{\sqrt 2e^{-3f/2}\over 12}\left(-dx^A\partial_Aw-
dx^\mu\Gamma_\mu \Gamma^A\partial_Aw+2dx^A\Gamma_{AB}\partial^Bw\right)\eta=
0.\cr}}
Setting successive terms to zero, one can readily determine $f$,
$b$, and $\psi$.  In particular, one gets
\eqn\yumbo{{1\over 8}\partial_Bf={\sqrt 2 e^{-3f/2}\over 6}\partial_Bw,}
so that
\eqn\zumbo{e^{3f/2} = c+ 2\sqrt 2 w }
with $c$ a constant.  
(In particular, $f$ is as promised above a function of $w$, so that
the differential form $G$ is uncorrected from the solution found with
the unperturbed metric.)
Moreover, one gets 
\eqn\tumbo{{1\over 8}\partial_Ab=-{e^{-3f/2}\sqrt 2\over 12} \partial_Aw
=-{\sqrt 2\over 12} {\partial_A w\over c+2\sqrt 2 w}.}
so that up to an irrelevant additive constant,
\eqn\yepto{b=-{1\over 3}\ln(c+2\sqrt 2 w).}
The metric \looko\ thus turns out to be
\eqn\impumbo{ds^2=(c+2\sqrt 2 w)^{-1/3}\eta_{\mu\nu}dx^\mu dx^\nu
+(c+2\sqrt 2 w)^{2/3}g_{AB}dx^Adx^B.}

As a check, this can be compared to the special case of the
eleven-dimensional extreme five-brane solution, as obtained by Guven
\ref\guven{R. Guven, Phys. Lett. {\bf B276} (1992) 49.}.
In this case, K is replaced by
 $\R^5$, and the ``unperturbed'' metric
on $\R^{11}=\R^6\times {\rm K}$ (before turning on the $G$ field) 
is simply the flat metric $\eta_{\mu\nu}dx^\mu dx^\nu
+\delta_{AB}dx^Adx^B$.  $G$ is then taken to be the magnetic
field due to a point charge at $x^A=0$, so that $w=q/R^3$,
where $R=\sqrt{x_Ax^A}$ and $q$ is the charge.   By scaling
one may set $c=1$.  If
we let
\eqn\delot{\Delta^{-1} =c+2\sqrt 2 w = 1+{2\sqrt 2 q\over R^3},}
then the metric \impumbo\ becomes
\eqn\yiedo{ds^2=\Delta^{1/3} \eta_{\mu\nu}dx^\mu dx^\nu+\Delta^{-2/3}
\delta_{AB}dx^Adx^B,}
which is a standard form of the five-brane solution.  Similarly,
one may take 
\eqn\udisp{w=\sum_i {q_i \over |\vec x - \vec x_i|^3}}
 and recover
a solution with parallel five-branes in $\R^{11}$.

Note that as long as $q>0$, which is the correct sign of the charge for
a five-brane that obeys the supersymmetry condition \yulu\
(anti-fivebranes with $q<0$ would require a supersymmetry
condition obtained by changing the sign on the right hand side of
\yulu), the function $\Delta^{-1}=c+2\sqrt 2 w$ is positive
definite throughout $\R^5-\{0\}$, so the metric is completely
sensible outside of the origin.  (It can in fact also be continued
past $R=0$ \ref\duffo{M. J. Duff, G. W. Gibbons, and P. K. Townsend,
Phys. Lett. {\bf B332} (1994) 321.}.)  
That is compatible
with the fact that (by considering a superposition of many parallel
five-branes that become coincident at the origin) one could
obtain arbitrarily big $q$, so that the description by the long
wavelength equations we have been using would be valid arbitrarily
close to $R=0$.  

An individual five-brane has a  charge quantum  $q=q_0$ with $q_0$  of
order $\kappa^{2/3}$.  For such an individual five-brane, the solution
given above in terms of classical supergravity is only valid at
$R>>\kappa^{2/9}$, that is, it breaks down at a Planck length
from $R=0$.

There is one important situation in which --  keeping the supersymmetry
condition precisely as in \yulu\ -- one does meet an object
very similar to a five-brane but with $q<0$.  This is the case
of a codimension five $\Z_2$ fixed point, which carries \otherwitten\
a five-brane charge $q=-q_0/2$.  
In this case, positivity fails  at $R$ of order $|q|^{1/3}$,
that is at $R$ of order the Planck length, and the description
breaks down there.    A macroscopic failure of the description
would occur if one could get a macroscopic negative $q$.
One cannot, however,
 use $\Z_2$ orbifold singularities to obtain a fivebrane
charge more negative than $-q_0/2$ except by making the singularities
meet, obtaining a worse singularity of space-time that would need
a separate analysis. 

Now we consider the other case in which the sources are codimension
one $\Z_2$ fixed points.  This arises for ${\rm K}={\rm K3}\times \S^1/
\Z_2$.  For brevity, we will omit the five-branes; their inclusion
would not greatly change things.

We start with a bare metric $ds^2_{(0)}$ on ${\rm K3}\times \S^1$
such that the K3 has volume $V_0$ and the $\S^1$ has circumference
$\rho_0$.  (More specifically,
take the metric on the $\S^1$ to be simply $(dx^{11})^2$ where
$x^{11}$ runs from $0$ to $2\pi \rho_0$.)
Then we determine $w$ by the equation $\nabla_0w={\rm
``sources}$.''  This only determines $w$ up to an additive constant.
To fix the constant, we may proceed as follows.  On the interior of
${\rm K3}\times \S^1/\Z_2$, $w$ is harmonic and so cannot have a minimum.
(This is still true if five-branes are present, as $w\to +\infty$ near
a five-brane.)
The minimum of $w$ is thus automatically 
at one of the $\Z_2$ fixed points, and there is no essential loss in
assuming that this is at $x^{11}=\pi\rho_0$.  By adding a constant
to $w$, one can suppose that $w$ vanishes at its minimum.     
The solution \impumbo\ for the exact metric is thus regular as long
as $c>0$, and develops a singularity on the boundary precisely at $c=0$.
This is the singularity encountered in \duff\ and also in section two of
the present paper; we would like to understand it a little better.

When the dimensionless number $\rho_0/V_0^{1/4}$ is of order one,
 by the time one reaches $c=0$,
the volume of ${\rm K3}\times \S^1/\Z_2$ is Planckian and our approximations
are no longer valid.  The interesting case to look at is
$\rho_0/V_0^{1/4}>>1$.  In this case, the qualitative behavior of $w$
is as follows.

Define a function of $x^{11}$ by 
\eqn\ikko{z(x^{11})=\int_{\rm K3}w \sqrt{ g_{(0)}} d^4x,}
with the integral taken at fixed $x^{11}$.  
The equation $\nabla_{(0)}w=0$ implies that
\eqn\pikko{{\partial^2 z\over \partial (x^{11})^2}=0}
so that $\partial z/\partial x^{11}$ is constant.
The value of the constant follows from \roaring:
\eqn\heffo{{\partial z\over \partial x^{11}}=-6 \pi\sqrt 2\left(\kappa\over
4\pi\right)^{2/3} (k-12),}
where $k$ is the instanton number at $x^{11}=0$.  (The instanton number
at $x^{11}=\pi \rho_0$ is $24-k$ so that the Bianchi identity has a solution.)
Thus for $k=12$, $z$ is constant and the volume of the K3 remains bounded
away from zero even for $\rho\to\infty$; this is the case studied in 
\duff, which leads to the most straightforward string-string duality.
We want to look at the case $k\not= 12$ where a singularity will develop
at finite $\rho$.  Up to a change of coordinates $x^{11}\to \pi\rho_0-
x^{11}$, we can assume that $k>12$; in fact, assuming as above that
the minimum of $w$ is at $x^{11}=\pi\rho_0$ (rather than $x^{11}=0$) is
equivalent to taking $k>12$.

The case in which our approximations say something interesting about the
singularity is the case in which $\rho_0^4/V_0>>1$.  In this limit,
K becomes a long tube, and one can to good approximation
ignore the dependence of $w$ on the
``small'' K3 directions, so 
$w$ is almost a function of $x^{11}$ only.  In that approximation,
the Laplace equation for $w$ reduces to $\partial^2w/\partial (x^{11})^2=0$,
with solution
\eqn\hujju{ w = {\pi\rho_0-x^{11}\over V_0}6\pi\sqrt 2\left(\kappa\over
4\pi\right)^{2/3}(k-12).}
The error in this formula is of order
\eqn\killo{ \bar w = {\kappa^{2/3}\over V_0^{3/4}}.}
(This is estimated as follows.
If one studies the equation for $w$ near $x^{11}=0$ or $x^{11}=\pi\rho$,
one sees that the deviation from \hujju\ is determined by
 a linear equation and
a boundary condition that are independent of $\rho$ for $\rho\to\infty$,
so the correction to \hujju\ is $\rho$-independent for large $\rho$;
\killo\ then follows by dimensional analysis given that the source
term in the equation is proportional to $\kappa^{2/3}$.) 
$\bar w$ can also serve as an estimate for the order of magnitude
of $w$ at $x^{11}=\pi\rho_0$.  

The volume $V$ of the K3 at $x^{11}=0$ (in the exact metric),
is 
\eqn\kluppo{V=V_0(c+2\sqrt 2w(x^{11}=0))^{4/3}.}
At the critical point, $c=0$, this volume is
\eqn\juppo{V=\left(24\pi^2
\left({\kappa\over 4\pi}\right)^{2/3}(k-12)\right)^{4/3}\left({\rho_0^4\over
V_0}\right)^{1/3}.}
The critical value of $\rho$ is $\rho={1\over \pi}\int_0^{\pi\rho_0}
dx^{11} e^{f/2}$ or 
\eqn\guffle{\rho={3\over 4}\left({\rho_0^4\over V_0}\right)^{1/3}
\left(24\pi^2\left({\kappa\over 4\pi}\right)^{2/3}(k-12)\right)^{1/3}.}
So the critical value of $V/\rho$ is
\eqn\piffle{{V\over \rho} = 32\pi^2(k-12)\left({\kappa\over 4\pi}
\right)^{2/3}.}

What characterizes the critical point is that, while the gauge coupling
is small in one $E_8$, it becomes strong in the second $E_8$.  In fact,
for sufficiently big $\rho_0^4/V_0$, the critical value of the volume
$V$ in \juppo\ can be as big as one wants, the coupling in the $E_8$
that is supported at $x^{11}=0$ then being of order $1/V$.  
On the other hand, using \killo, the volume in the second $E_8$
at the critical point is of order $\bar w^{4/3} V_0=\kappa^{8/9}$,
that is of order the Planck volume, independent of $\rho_0/V_0$.
The gauge coupling in the second $E_8$ is thus apparently of order
one at the critical point.

In \duff, it appeared that the gauge coupling in the second $E_8$ is infinite
at the critical point, corresponding to the K3 at $x^{11}=\pi \rho_0$ having
zero volume there.  In the present approach, the volume appears to be
Planckian rather than zero (and therefore the gauge coupling appears to
be of order one rather than infinite).  But as our approximations in the
present discussion break down in any case
when the volume is Planckian, it does not
seem that we have any real evidence for a phase transition occurring
while the gauge coupling is still finite.

\listrefs
\end